\documentclass[sigconf,authorversion]{acmart}

\usepackage{listings}
\AtBeginDocument{%
  \providecommand\BibTeX{{%
    \normalfont B\kern-0.5em{\scshape i\kern-0.25em b}\kern-0.8em\TeX}}}

\copyrightyear{2022}
\acmYear{2022}
\setcopyright{acmcopyright}\acmConference[SBST'22 ]{The 15th Search-Based Software Testing Workshop}{May 9, 2022}{Pittsburgh, PA, USA}
\acmBooktitle{The 15th Search-Based Software Testing Workshop (SBST'22 ), May 9, 2022, Pittsburgh, PA, USA}
\acmPrice{15.00}
\acmDOI{10.1145/3526072.3527537}
\acmISBN{978-1-4503-9318-8/22/05}

\begin{document}

\title{Towards Run-Time Search for Real-World Multi-Agent Systems}


\author{Abigail C. Diller}
\affiliation{%
  \institution{Grand Valley State University}
  \streetaddress{1 Campus Dr.}
  \city{Allendale}
  \state{Michigan}
  \country{USA}
  \postcode{49401}
}
\email{dillerab@mail.gvsu.edu}

\author{Erik M. Fredericks}

\affiliation{%
  \institution{Grand Valley State University}
  \streetaddress{1 Campus Dr.}
  \city{Allendale}
  \state{Michigan}
  \country{USA}
  \postcode{49401}
}

\email{frederer@gvsu.edu}

\renewcommand{\shortauthors}{Diller and Fredericks.}

\begin{abstract}
Multi-agent systems (MAS) may encounter uncertainties in the form of unexpected environmental conditions, sub-optimal system configurations, and unplanned interactions between autonomous agents. 
The number of combinations of such uncertainties may be innumerable, however run-time testing may reduce the issues impacting such a system.
We posit that search heuristics can augment a run-time testing process, in-situ, for a MAS.  To support our position we discuss our in-progress experimental testbed to realize this goal and highlight challenges we anticipate for this domain.

\end{abstract}



\maketitle

\section{Introduction}

Run-time testing can provide assurance that a multi-agent system (MAS) will continuously satisfy requirements and behave as expected even when faced with uncertainty~\cite{muccini.2016}. 
However, MAS testing is a challenge as it must account for the unique characteristics of heterogeneous agents and may require different testing approaches~\cite{nguyen.2007,nguyen.2012}. 
Furthermore, the number of potential combinations of system conditions, environmental interactions, and network issues that a MAS may experience during its life cycle may be impossible to enumerate~\cite{yong.2019}.
We discuss the application of search-based software testing (SBST) techniques as an approach for navigating the complex solution space of MAS validation/verification tasks.
We next present our work-in-progress experimental testbed, outline key challenges for the MAS domain, and highlight relevant research throughout the paper.


\section{GreenRoom: Experimental Testbed}

A MAS is typically modeled as a network of agents (i.e., devices or simulated processes) that each act independently in support of their own goals while maintaining a common set of shared objectives~\cite{kolp.2002,bresciani.2004.tropos}.  Agents monitor the operating context of both the system and environment and autonomously fulfill both local (i.e. impacting the agent itself) and global (i.e., impacting the system) objectives.  Additionally, agents will generally communicate with other agents via message passing to ensure that knowledge is shared.  

To support MAS execution, run-time requirements monitoring~\cite{sawyer.2010}, run-time testing~\cite{fredericks.2013.SEAMS}, and run-time verification~\cite{ghezzi.2010, lim.2016} have been proposed as techniques for ensuring the system continuously satisfies key objectives (i.e., to ensure the MAS continuously expresses acceptable behaviors per its artifacts).  Moreover, genetic algorithms~\cite{holland.1992} have been previously applied for hardware-based optimization in-situ for field-programmable gate arrays~\cite{peker.2018} as well as general search heuristics on constrained devices~\cite{fredericks.2020.ssbse}. 
Similar to cyber-physical systems~\cite{fredericks.2019.sbst}, a MAS can experience complex real-world concerns such as power consumption, human interaction, and unexpected weather conditions.
Most similarly, Nguyen \textit{et al.} explored evolutionary approaches for continuous MAS evolutionary testing~\cite{nguyen.2007,nguyen.2012}, however we focus on hardware applications in-situ and the related concerns therein (e.g., timing delays, lost/corrupted data, etc.). 

\begin{figure}[htb]
    \centering
    \includegraphics[width=2.2in]{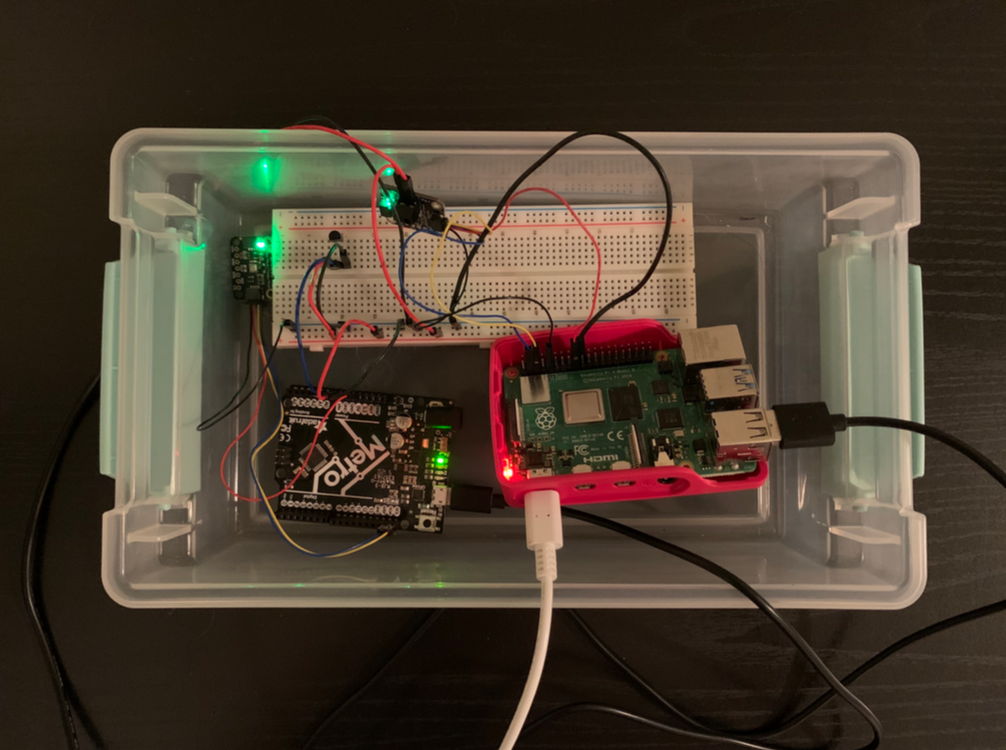}
    \caption{Experimental testbed comprising an Arduino Uno, Raspberry Pi 4, and wired sensors.}
    \label{fig:bench}
\end{figure}

Figure~\ref{fig:bench} shows \texttt{GreenRoom}, our hardware-based experimental testbed that both implements and extends our prior work~\cite{fredericks.2019.sbst}.
We model our system as a MAS, where one agent (i.e., an Arduino Uno) is tasked with ensuring that the local environmental conditions are within an acceptable range (e.g., comfortable temperature, no dangerous gasses, etc.) and another agent (i.e., a Raspberry Pi 4) is tasked with continuously generating test data via a genetic algorithm~\cite{holland.1992} as well as monitoring power consumption of the system as a whole. Agents communicate via a serial connection to transmit/receive data as needed.   

We next outline our procedure for performing SBST at run-time, comprising a \textit{testing agent} (i.e., Raspberry Pi) and a set of \textit{agents under test} (i.e., Arduino or other devices):

\begin{enumerate}
    \item \textit{Testing agent} configures search procedure at start of execution and continuously generates sets of test data relevant to the \textit{current conditions} experienced by the MAS.
    \item \textit{Testing agent} packages test data and disseminates to target \textit{agents under test}.
    \item Each \textit{agent under test} parses received test data, executes test case(s), and transmits results back to \textit{testing agent}.
    \item \textit{Testing agent} collates test results, incorporates into fitness function (see Equation~\ref{eqn:ff}), and executes the subsequent generation of search.
\end{enumerate}

While each search problem is different, we anticipate that a weighted fitness function can provide a lightweight method for guiding the search procedure.  For \texttt{GreenRoom}, our proposed fitness function (i.e., Equation~\ref{eqn:ff}) comprises \textit{maximizing} the number of test case failures and \textit{maximizing} novelty of test data while intelligently exploring the solution space (i.e., via novelty search~\cite{lehman.2011}).  For the purpose of this paper, we abstract the sub-fitness functions for maximizing test failures and test data novelty, respectively.

\vspace{-0.2in}

\begin{equation}
   {ff} = \alpha_{fail} * max({TC}_{fail}) + \alpha_{novelty} * max({novelty})
    \label{eqn:ff}
\end{equation}

Equation~\ref{eqn:tc} presents a sample test case that checks the validity of monitored temperature sensor readings as well as test data generated by our testbed.  At present, our Raspberry Pi is using EasyGA\footnote{See \url{https://pypi.org/project/EasyGA/}.} to generate test data, a serial connection between the Raspberry Pi and Arduino to communicate, and injected sensor values to test the rigor of our bench setup. 

\vspace{-0.15in}

\begin{equation}
  {TC_{temp}}(t) = (t \geq -40{~}\&\&{~}t \leq 85.0){~}?{~}{true}{~}:{~}{false}
    \label{eqn:tc}
\end{equation}


Our genome comprises a list of values that reflect the input data of each test case. There are $20$ test case templates currently implemented in \texttt{GreenRoom} leading to a small search space.
We next highlight challenges we have faced thus far in applying SBST to our experimental testbed.



\noindent \textbf{Challenge (1) - Search In-Situ}: Testing a MAS during execution in its deployed environment has the benefit of using real-world data at the detriment of potentially impacting the system under test.  As such, care must be taken to avoid disruption of critical system features/behaviors (e.g., system invariants).

\noindent \textbf{Challenge (2) - Navigating the Search Space at Run Time}: The search space must be efficiently explored with minimal resources to avoid impacting ``normal'' system operations.  A lightweight search heuristic can enable online search (e.g., a (1+1) evolutionary algorithm~\cite{bredeche.2010}) to the detriment of searching power (e.g., local optima, premature convergence, etc.).  As such, a fine balance between search capabilities and run-time impact must be considered.

\noindent \textbf{Challenge (3) - Managing System Constraints}: Similar to Challenge (2), agents within a MAS may be constrained devices that operate with minimal memory, storage, and power capabilities.  For example, communication via serial connection between the Arduino and Raspberry Pi must follow strict timing constraints to avoid lost or garbled data.  System capabilities and constraints must be considered when overlaying run-time testing techniques to avoid negatively impacting the agent(s) under test.

\section{Discussion}

This paper has discussed early-stage research for \texttt{GreenRoom}, our experimental testbed for performing run-time SBST on a system that has been modeled as a MAS.  At present, we have a preliminary hardware setup comprising multiple agents with individual goals, including generation of test data during execution and communication of test data/results between agents.  We have also identified challenges that can be experienced by researchers in this domain.  Future work for this project involves significant extension to our body of test cases, incorporation of power monitoring metrics into our search procedure, and a full empirical evaluation.

\begin{acks}
This work has been supported in part by grants from the Michigan Space Grant Consortium (\#80NSSC20M0124) and Grand Valley State University.  The views and conclusions contained herein are those of the authors and do not necessarily represent the opinions of the sponsors.
\end{acks}

\bibliographystyle{ACM-Reference-Format}
\bibliography{efredericks_master}

\end{document}